\documentclass{iau} 
\usepackage{graphicx,cite}

\title[Commissioning of FLAG] 
{Commissioning of FLAG: A phased array feed for the GBT}

\author[K. M. Rajwade et el.]   
{K. M. Rajwade$^{1,2}$,
 N. M. Pingel$^1$, R. A. Black$^3$, M. Ruzindana$^3$, M. Burnett $^3$, B. Jeffs$^3$, K. Warnick$^3$, D. J. Pisano$^1$, D. R. Lorimer$^1$, R. M. Prestage$^4$, L. Hawkins$^4$, J. Ray$^4$, P. Marganian$^4$, T. Chamberlin$^4$, J. Ford$^4,6$, W. Shillue$^5$, D. A. Roshi$^5$}
 
\affiliation{$^1$Department of Physics and Astronomy, West Virginia University
                 Morgantown, WV 26505, USA\\ email: {\tt kaustubh.rajwade@manchester.ac.uk} \\[\affilskip]
$^2$Jodrell Bank Centre for Astrophysics, University of Manchester, Oxford Road, Manchester M193PL, UK \\[\affilskip]
$^3$ Department of Electrical Engineering, Brigham Young University, Provo, UT, USA \\[\affilskip]
$^4$ Green Bank Observatory, Green Bank, WV, USA \\[\affilskip]
$^5$ National Radio Astronomy Observatory, CDL, Charlottesville, VA, USA\\
$^6$ Steward Observatory, University of Arizona, Tuscon, AZ, USA}

\pubyear{2017}
\volume{337}  
\jname{Pulsar Astrophysics -­ The Next 50 Years}
\editors{P. Weltevrede, B.B.P. Perera, L. Levin Preston \& S. Sanidas, eds.}

\begin{document}

\maketitle

\begin{abstract}
Phased Array Feed (PAF) technology is the next major advancement in radio astronomy in terms of combining high sensitivity and large field of view. The Focal L-band Array for the Green Bank Telescope (FLAG) is one of the most sensitive PAFs developed so far. It consists of 19 dual-polarization elements mounted on a prime focus dewar resulting in seven beams on the sky. Its unprecedented system temperature of$\sim$17~K will lead to a 3 fold increase in pulsar survey speeds as compared to contemporary single pixel feeds. Early science observations were conducted in a recently concluded commissioning phase of the FLAG where we clearly demonstrated its science capabilities. We observed a selection of normal and millisecond pulsars and detected giant pulses from PSR B1937+21.
\end{abstract}

\firstsection 
\section{Introduction}
\begin{figure}[ht]
\begin{center}
 \includegraphics[scale=0.11]{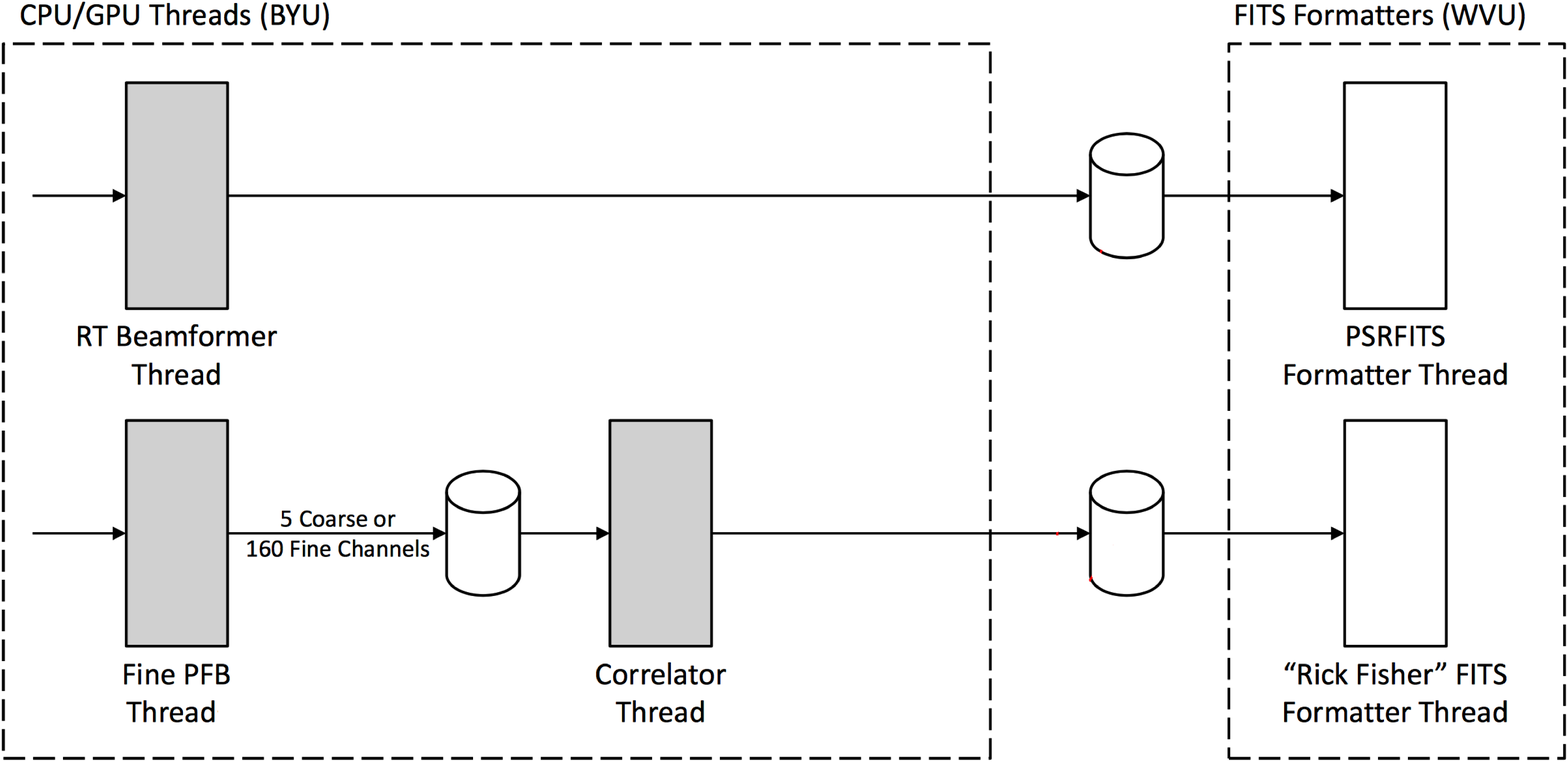}
 \caption{Block diagram of the FLAG backend. The schematic shows the processing pipeline for the correlator/beamformer in one GPU. The beamformer and the correlator threads are handled by the GPU while the FITS writer thread is handled by the CPU.}
   \label{fig0}
\end{center}
\end{figure}

PAF technology is being developed for most state of the art radio telescopes (Chippendale et al. 2010). These receivers enable faster survey speeds for pulsar and fast transient searches. The Green Bank Telescope (GBT), with its excellent aperture efficiency and large area, is an ideal telescope to 
build a PAF for. The FLAG was developed in a joint collaboration between the Green bank Observatory (GBO), National Radio Astronomy Observatory (NRAO), Brigham Young University (BYU) and the West Virginia University (WVU). The backend consists of five high performance computers (HPCs), each comprising two Nvidia GTX780 Ti Graphical Processing Units (GPUs) to perform real-time beamforming. Each GPU handles one-twentieth of the band with non-contiguous channels to give a total processing bandwidth of 150 MHz (see Figure.~\ref{fig0}). The stitching of the band and final data products are processed offline. The first pulsar observations were conducted in May of this year where three pulsars, namely PSR B1933+16, PSR B1929+10, and PSR B1937+21 were observed in the central beam. The second set of observations were conducted in July of this year where PSR B0329+54 was observed in all seven beams for which the data are still being processed.

\section{Results}
\begin{figure}[ht]
\begin{center}
 \includegraphics[width=60mm,height=46mm]{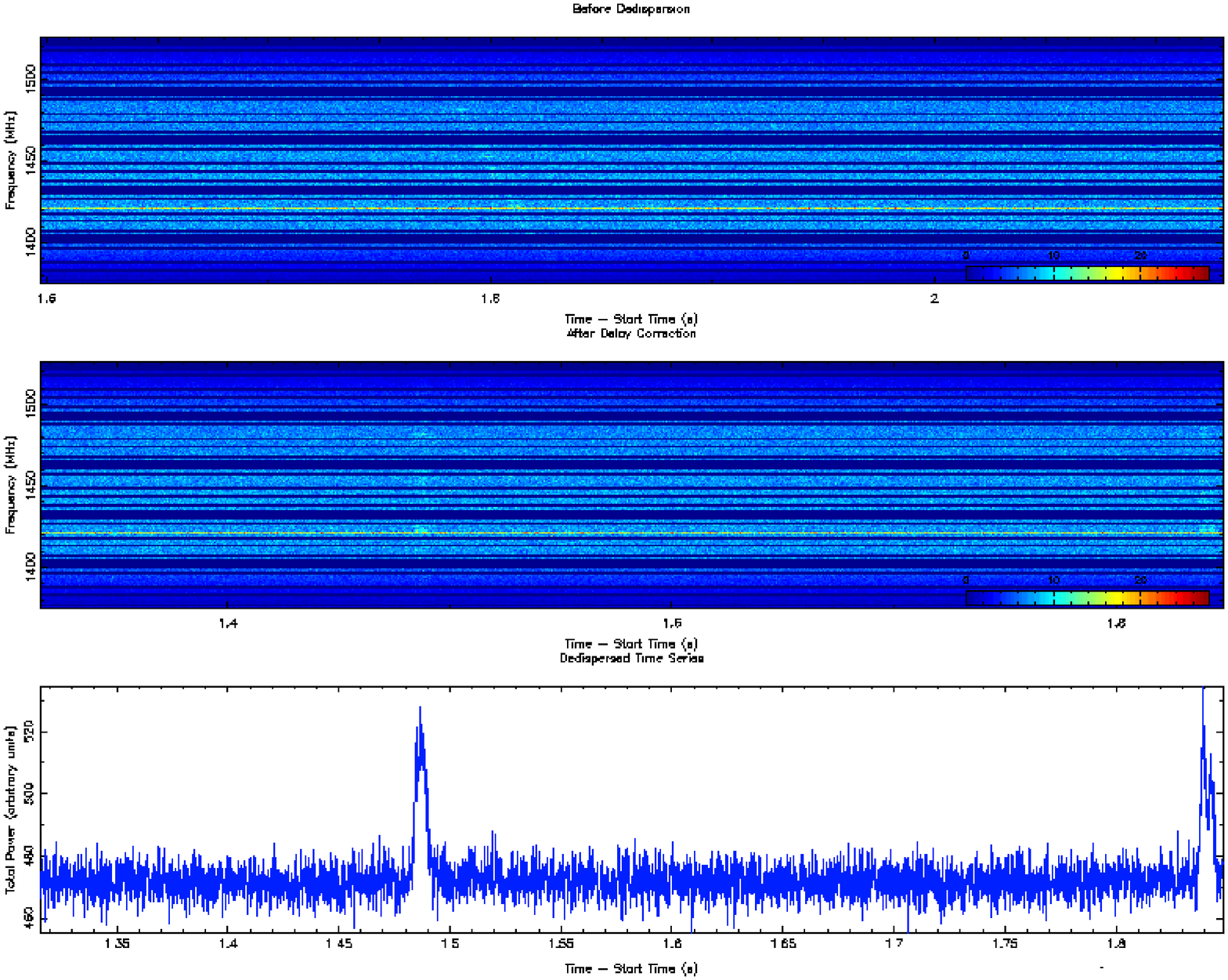}
 \includegraphics[width=60mm,height=47mm]{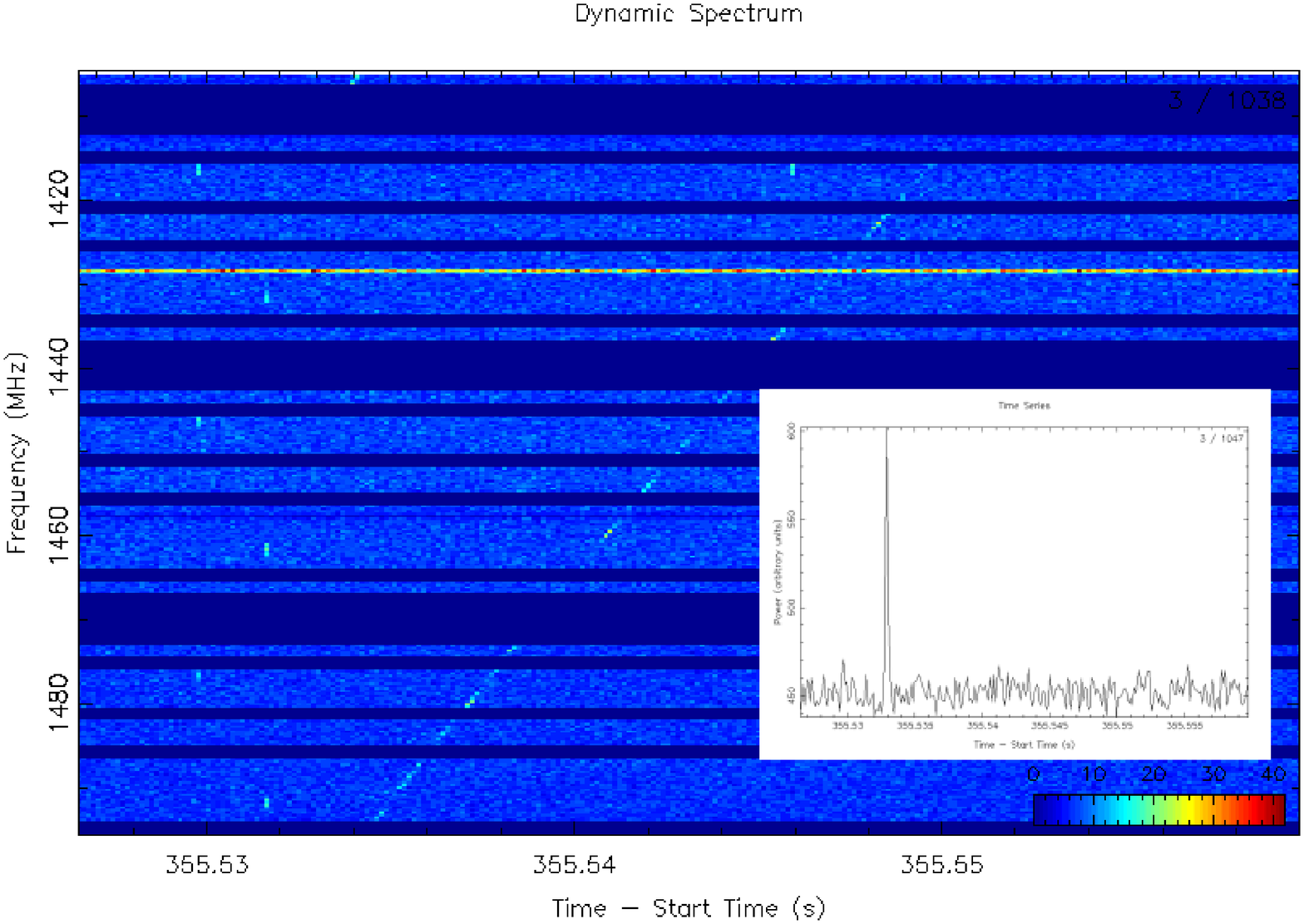}
 \caption{Left Panel shows single pulses from PSR B1933+16 along with the dynamic spectrum. Right Panel shows dynamic spectrum of a giant pulse of PSR B1937+21. The inset shows the frequency collapsed time series for the same.}
   \label{fig1}
\end{center}
\end{figure}

We detected single pulses from PSR B1933+16 as shown in figure~\ref{fig1} which demonstrated the excellent sensitivity of the PAF. Moreover, we observed giant pulses from PSR B1937+21 (Right panel of figure.~\ref{fig1}). We detected PSR B1929+10 in the periodicity search though we did not observe single pulses from the same. This was unexpected given the flux density of the pulsar (S$_{1.4}\sim$36~mJy). The low signal to noise detection was attributed to the modulation in the observed flux due to strong scintillation effects (Wang et al. 2005). The results, when compared with the sensitivity of the single pixel receiver at the GBT with the GUPPI backend (800~MHz bandwidth), show that we obtain an improvement in the survey speeds by a factor of $\sim$ 3.1.

\end{document}